\documentclass[aps,showpacs,prb,twocolumn,supperscriptaddress]{revtex4-1}
\usepackage{graphicx} 
\usepackage{dcolumn} 
\usepackage{bm} 
\usepackage{graphicx,color,epsfig,rotate}
\usepackage{amsmath,amssymb}

\begin{document}
\title{Understanding glass-like  Vogel-Fulcher-Tammann equilibration times:\\ 
 microcanonical effective temperatures  in  quenched 3D  martensites}
\author{N. Shankaraiah${^1}$, K.P.N. Murthy${^2}$ and S.R. Shenoy${^1}$ }
\affiliation{ ${^1}$Tata Institute of Fundamental Research-Hyderabad, Hyderabad, Telangana 500046, India. \\
${^2}$Dept of Physics, Central University of Rajasthan, Bandar Sindri, Rajasthan 305817, India}

\date{\today}

\begin{abstract}

\noindent We do Monte Carlo simulations of four 3D structural transitions, with vector-spin models of their martensitic strain domains under quenches to $T$, to test a generic post-quench Partial Equilibration Scenario (PES) of Ritort. We  indeed confirm that energy-lowering passages between fixed-energy shells induce a signature PES distribution of an exponential tail in  heat releases, scaled in an effective search temperature. A linear vanishing of this $T_{eff}(T)\sim T_d -T$  at a temperature $T_d$ where PES passage-searches freeze,  explains the   Vogel-Fulcher like  divergence of equilibration times $ e^{1/ T_{eff} (T)}\sim e^{1/(T_d -T)}$, extracted from  incubation-time delays of simulations and martensitic alloys.

 \end{abstract}  
 
\maketitle

\vskip 0.5truecm

Glassy freezing or structural arrest of a rapidly cooled liquid or colloidal system \cite{R1,R2,R3,R4,R5}  that pre-empts crystallisation, has been investigated for more than a century. Supercooled liquid models can yield heterogeneous domains of competing crystal structures\cite{R3,R5}.  Equilibration time  divergences at  a glassy freezing temperature $T_G$,  have been fitted to Vogel-Fulcher-Tammann  (VFT)  $\sim e^{1/T- T_G}$, or other forms \cite{R2,R4}. It is natural to study generic equilibration scenarios \cite{R6,R7,R8,R9} in specific structural-domain systems that have long relaxation times \cite{R1,R2,R3,R4,R5,R10,R11,R12,R13,R14,R15,R16}.

After a sudden quench, a system on  a  free energy landscape, has competing  pathways  to the new global minimum, delayed by free energy barriers  $\{\Delta F =\Delta U - T \Delta S\}$. The delay rates $e^{-\Delta F/T}$ will be from energy barriers ($\sim e^{-\Delta U/T}) $ and entropy barriers ($\sim e^{-|\Delta S|}$), schematically depicted in Fig 1.  Ritort and colleagues \cite{R6,R7,R8,R9} have  proposed a Partial Equilibration Scenario  (PES) for re-equilibrations delayed by {\it entropy} barriers.  Over a waiting time $t_w$, a  post-quench ageing system rapidly explores configuration shells of energy $ E(t_w)$, entropy $S (E)$, and (inverse) micro-canonical effective temperature  $1/T_{eff} (t_w)\equiv dS (E)/dE $. Passages to a lower shell of  $E(t_w +1) \equiv  E' =E(t_w ) +\delta E$ are driven by  spontaneous heat  releases ($\delta E =\delta Q < 0$)  to the bath at $T$. The PES says that an  iteration of  these cooling  steps ratchets  the system down to the new canonical equilibrium.
 The non-equilibrium probability  distribution for energy changes  \cite{R6,R8}  $P_0(\delta E; t_w)$ is peaked at positive energies, with an  exponential tail for $\delta E < 0$,  whose fall-off  $\sim e^{ \delta E/2 T_{eff} (t_w)}$ determines the effective temperature. The PES  distribution has been studied  by  {\it analytic} Monte Carlo (MC) methods  for harmonic oscillators \cite{R8}, and by numerical MC  simulations of  spin glasses and Lennard-Jones liquids \cite{R7,R9}.
We note that if the  effective temperature of the heat-release probability vanishes at some $T=T_d$,  then there is an arrest  of the PES cooling process.

 \begin{figure}[h]
\begin{center}
\includegraphics[height=4.0cm, width=8.5cm]{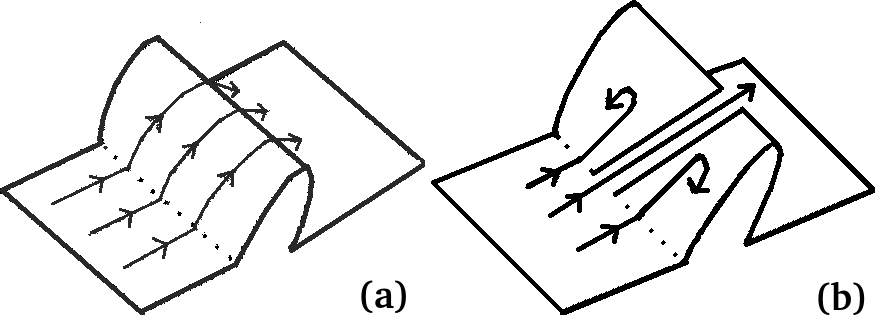}
\caption{{ \it  Schematic of delays from two limits of free energy barriers:}  a) Energy-barrier delays from thermally activated jump attempts. b) Entropy-barrier delays from searches for rare passages. Key seeks lock, most attempts fail. }
\end{center}
\end{figure}

 We consider solid-solid structural transitions of martensites \cite{R10,R11,R12,R13,R14,R15,R16}, quenched below a thermodynamic $T_0$, with competing   domains  and  slow relaxations \cite{R11,R17}. Martensites undergo first-order, diffusionless  transitions \cite{R5,R10} from the higher-symmetry austenite, with atomic shifts locked to  their unit-cell distortions  (`military transformations'). The order-parameter strains have degenerate lower-symmetry `variants' separated by crystallographically oriented Domain Walls (DW), that can  form complex microstructures \cite{R10,R14,R15}. A long-standing puzzle \cite{R11,R12} is that while quenches  of austenite to below an (athermal) `martensite start'  temperature $T_1 >T$  results in avalanche martensitic conversions,  quenches above it $T_1 < T$, show {\it delayed conversions} instead  of no conversions. Resistivity, as a transition diagnostic, is flat during  post-quench `incubations', that end in sudden drops at a delayed avalanche\cite{R11}. Delays rise sharply, for shallower quenches approaching a third temperature $T_d$ that is  in between, $T_1 < T_d < T_0$.  When a bath quench $T$ goes a few percent closer to $T_d$,  the resistivity-drops go from a few seconds after,  to ten thousand seconds  after, the temperature quench \cite{R12}. 
 
In this Letter we do MC simulations in  {\it three} dimensions,  with vector order parameter strains, for four structural transitions \cite{R18}.  We present here the cubic-tetragonal (CT) transition\cite{R15}, with  a strain order parameter of components  $N_{OP}=2$, with three competing unit-cell `variants' $N_V=3$. We confirm for all four transitions, that the PES energy change distribution has the predicted generic behaviour: an exponential tail, with an effective temperature that regulates heat releases\cite{R6,R7,R8,R9}.
 
 For our specific case of  quenches across a first order transition, the Order Parameter (OP) rises from zero,  enabling the waiting time $t_w$  to be defined by rising-OP marker events at $t_m$, that depend on $T$. This choice  $ t_w= t_m (T)$ induces quench-temperature dependences: $T_{eff} (t_w) \rightarrow T_{eff} (T)$ and  $P_0(\delta E;  t_w) \rightarrow P_0 (\delta E, T)$. For passages to lower energy shells, the OP evolution must satisfy  $T$-controlled entropy-barrier constraints, postulated as of two types:  a) A constraint that OP configurations must find and  enter a  Fourier space bottleneck that is like a  Golf Hole  (GH) that funnels into fast passage, as suggested for protein folding  \cite{R19};  or b) A constraint that  the OP states need transient catalysts to enable  fast passages, as inspired by facilitation models\cite{R20,R21,R22}.  Our case is a),  and we find a linear vanishing $T_{eff} (T)  \sim (T_d-T)$. The `search freezing' temperature $T_d$ occurs at a pinch-off  on warming, of  the  $\vec k$-space  {\it inner} radius of an angularly modulated bottleneck. Equilibration times are exponential in entropy barriers \cite{R11,R12,R23}, and for quenches   $T _d >T> T_1$,   diverge as $ {\bar t}_m (T)\sim e^{1/T_{eff} (T)} \sim e^{1/(T_d-T)}$. Thus VFT -like behaviour is not restricted to the glass transition. Conversely, entropy barriers vanish and delay times collapse for $T< T_1$, when the bottleneck  expands on cooling to span the Brillouin zone.

\begin{figure}[h]
\begin{center}
\includegraphics[height=4.0cm, width=8.5cm]{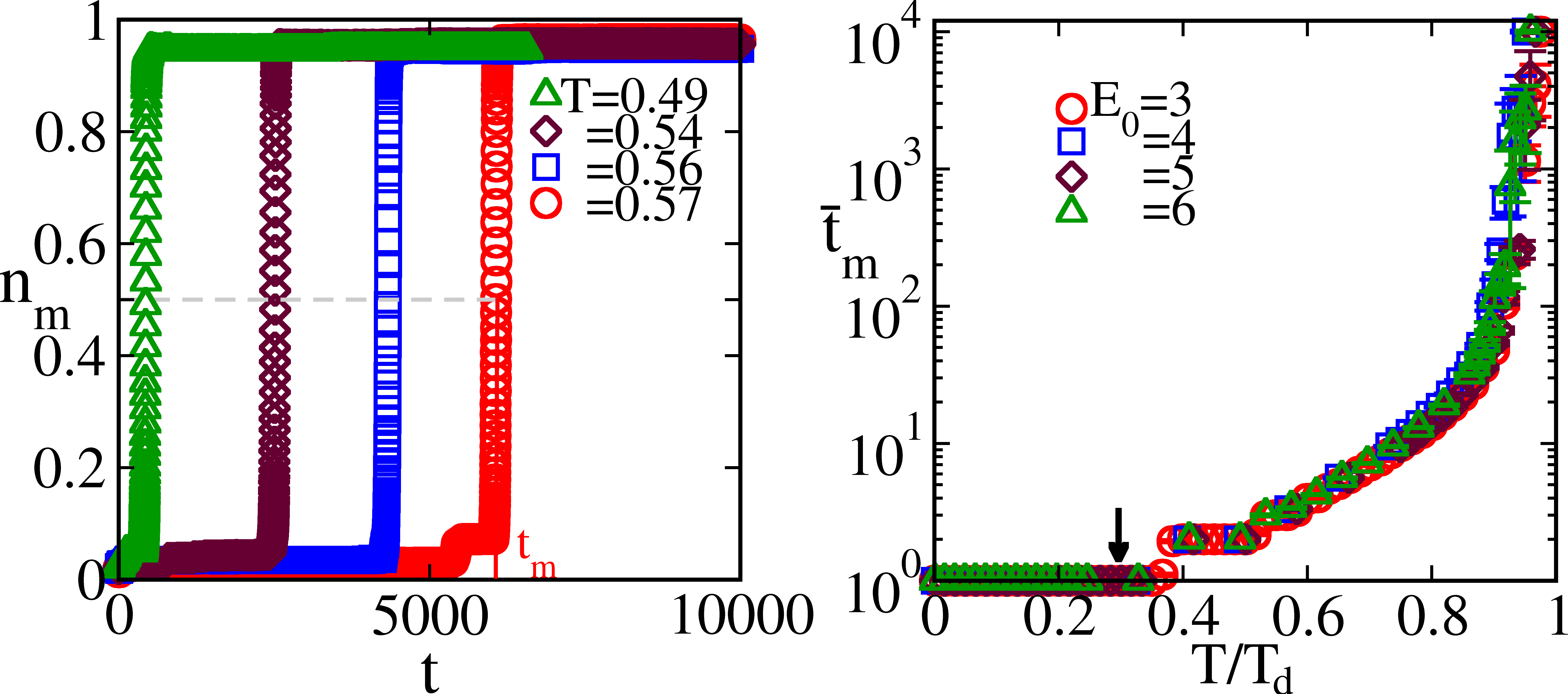}
\caption{{\it Delay times for  CT martensitic conversion: } The martensite fraction $n_m (t_m)=0.5$ defines $t_m$. a) For $T\leq T_1$  avalanche conversions occur,  at  $t_m =1$. For $T_d > T > T_1$   DW sluggishness causes `incubation'  delays or postponement of  conversion avalanches to  $t=t_m (T)$.  b) Log-linear plot of  mean delay time ${\bar t}_m (T)$ versus ${T/T_d} < 1$. Delay times  are  not  exponentially sensitive to Hamiltonian energy scales $E_0$, so are not activated:  delays are from {\it  entropy} barriers.}
\end{center}
\end{figure}  

 \begin{figure}[h]
\begin{center}
\includegraphics[height=4.0cm, width=8.5cm]{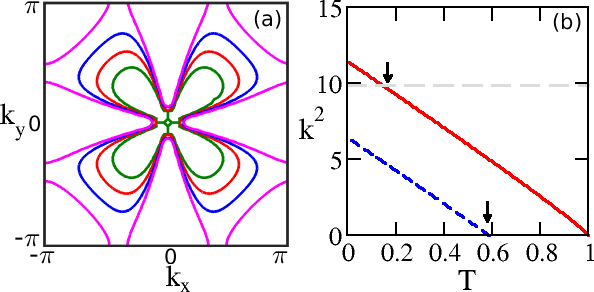}
\caption{{ \it Bottlenecks in Fourier space for CT transition: } The temperature dependence of bottleneck size and shape is shown for a [1,1,1] slice of  a 3D anisotropic bottleneck. a) The 2D slice in $(k_x, k_y)$ is like an anisotropic `Golf Hole', enclosing negative martensite states, that shrinks  with warming  $T$.  The open butterfly shape  changes topology to  a segmented four-petal flower shape at $T=T_d$. b) The anisotropic bottleneck  inner and outer radii $k_{in}, k_{out}$ are plotted as  $ {k}^2$   vs $T$. The bottleneck  outer radius $k_{out} (T)$ for   $0 < T \leq T_1$ (arrow) spans a Brillouin Zone size  of $\sim \pi$ (horizontal  light dashes),  so conversions are immediate.  The outer radius shrinks to a point on the right at the thermodynamic transition $T=T_0 =1$  to austenite-only states.  On the other hand, the inner radius $k_{in} (T)$ shrinks  on warming for   $0 \leq T \leq T_d$,  vanishing  (arrow) at   $ T=T_d $  when the outer radius is still nonzero: the bottleneck topology changes.
 }
\end{center}
\end{figure}

We  derive a  discretized-strain Hamiltonian \cite{R15}  in 3D,   from a crystal-symmetry invariant strain  free energy $F$,  that has Compatibility \cite{R14}, Ginzburg, and Landau terms in  $F/E_0 = \sum_{\vec r, \vec r'}  f_C +\sum_{\vec r}[ f_G + f_L]$, with $E_0$ an energy scale. There are six independent  {\it physical} strains\cite{R15} in 3D, that are linear combinations of Cartesian tensor strains: compressional $e_1$;  deviatoric or rectangular $e_2, e_3$ , and shear $e_4, e_5, e_6$.  The OP of the cubic-tetragonal (CT)  transition are two deviatoric strains  ${\vec e} =(e_3, e_2) =( \frac{1}{\sqrt{6}} \{e_{xx} +e_{yy}-2e_{zz}\}, \frac{1}{\sqrt{2}}\{e_{xx} -e_{yy}\})$. Austenite is  $\vec e =\vec 0$.  

The remaining $6-N_{OP}$  non-OP strains (one compressional and three shears) enter the Hamiltonian as harmonic springs. These are  minimized subject to a linear St Venant  Compatibility constraint\cite{R14}  that says no dislocations  are generated: the double-curl of the strain tensor must vanish. There are three independent algebraic equations in  $\vec k$ space, connecting  OP and non-OP strains \cite{R15}. The harmonic  non-OP strains then  analytically yield  an OP-OP interaction, whose transition-specific, {\it anisotropic} Compatibility kernel \cite{R15}  is a $2 \times 2 $matrix, $U_{\ell  \ell'} ({\hat k} )$ where $\ell, \ell' =2,3$. There is a prefactor of  $(1- \delta_{\vec k, 0})$, and dependence on  direction $\hat k ={\vec k}/|\vec k|$.

The Landau free energy for CT is  $f_L (\vec e) =[(\tau -1) {\vec e}^2 - 2(e_3 ^3 -3  e_3 e_2 ^2) + {\vec e}^{~4}]$ and has 4 minima, at $N_V= 3$ variants plus at zero strain. Here $\tau (T) \equiv (T- T_c)/ (T_0 -T_c)$, and $\tau (T_c) =0$  at the spinodal $T_c$, while $\tau (T_0)=1$ at the first-order transition temperature, scaled to be unity $T_0 =1$.   

In `polar' coordinates, $\vec e \equiv |\vec e|  \vec S$. Here the unit-magnitude `variant vectors' $\vec S (\vec r)$ specify the unit-cell  variants on either side of a Domain Wall (DW), that can be  martensite-martensite or martensite-austenite. The nonzero $N_V=3$ martensite-variants have spins \cite{R15}   $\vec S=  (S_3,S_2) =(1,0), (-1/2,{\sqrt{3} }/2), (-1/2,-{\sqrt{3} }/2)$, pointing to corners of an equilateral triangle in a unit circle, while  the  centroid $\vec S= (0,0)$ is  austenite. Thus ${\vec S}^2 =0$ or $1$.

The degenerate Landau minima are at mean-OP  magnitudes  $\bar \varepsilon (T) = (3/4) [1+\sqrt{1-(8 \tau/9)}]$. The variant domains  have mostly-flat strain magnitudes,  approximated  by $\bar \varepsilon (T)$. Substituting  ${\vec e} (\vec r)  \rightarrow  \bar \varepsilon (T) {\vec S}(\vec r)$, the Landau term becomes  $f_L (\vec e) \rightarrow  f_L (T) {\vec S} (\vec r)^2$. Here $f_L(T) \equiv  {{\bar \varepsilon} (T)}^2 g_L (T) \leq 0$, where $g_L = (\tau -1) + ( {\bar \varepsilon }(T) -1)^2  \leq 0$. At $T={T_0}^{-}$, the OP is unity $\bar \varepsilon=1$ and $g_L=0$.

 Notice a separation of time scale responses to $T$ quenches:  the OP magnitude $\bar \varepsilon (T)$  responds immediately, at $t=1$, while Domain Walls can take thousands of time steps $t$, to evolve successively  from DW Vapour to DW Liquid to a DW Crystal of twins. For our case of shallow quenches  $T_d > T > T_1$, it is the DW Vapour-to-Liquid conversion has  the long (bottleneck type) delays studied here. See Videos\cite{R21} A,B. The DW moves by correlated flips of spins that bracket it, while domain spins remain locked:  a  dynamical heterogeneity in space and time \cite{R3,R5}. ~[For deeper quenches $T<< T_1$ not studied here,  it is  the  DW Liquid-to-Crystal  twin orientation  that has long (facilitation type) delays. See Video\cite{R21} C.]

\begin{figure}[h]
\begin{center}
\includegraphics[height=4.0cm, width=8.5cm]{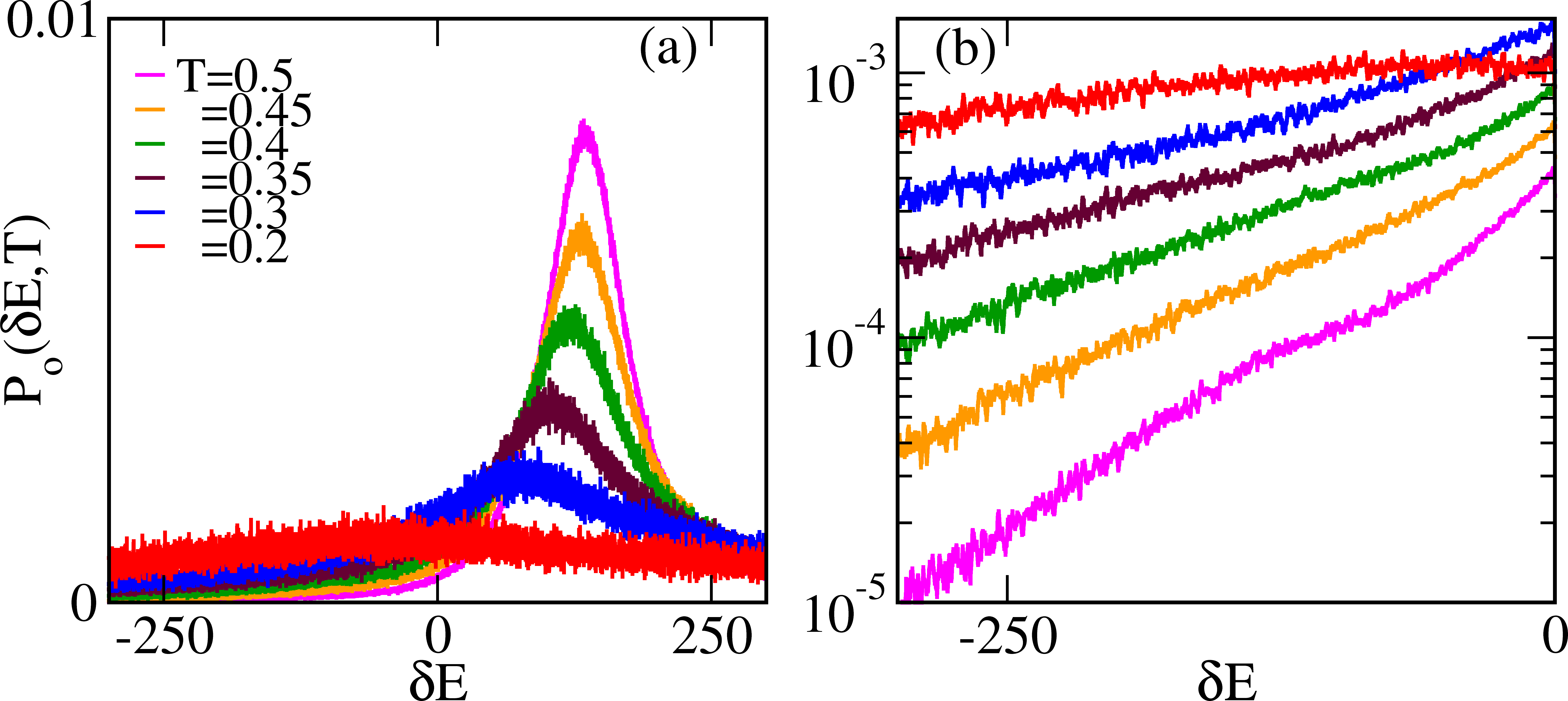}
\caption{{\it Energy-change  statistics for CT transition:}  a) Linear-linear plot of the normalized  probability $P_0(\delta E,T)$ versus energy change $\delta E$, for six $T$ quenches. b) Log-linear version. Slope at the origin $\beta_{eff}  (T)/2$ rises from zero for  $T > T_1$.  }
\end{center}
\end{figure}

The total hamiltonian is $\beta H = \beta H_L + \beta H_G + \beta H_C$, without extrinsic disorder. It  is diagonal in Fourier space, 
\begin{equation}
\begin{array}{rr}
\beta H = \dfrac{D_0}{2} [  \sum_{\ell, \ell'}  \sum_{\vec{k}}  [\epsilon_{\ell,\ell'} (\vec{k},T){\vec S}_{\ell}(\vec{k}) {\vec S}^*_{\ell'}(\vec{k} )] ,
\end{array}
\end{equation}  
with $D_0 \equiv 2 {\bar \varepsilon} (T)^2 E_0/ T$. The spectrum, with $K_\mu (\vec k) \equiv 2 \sin (k_\mu /2)$ and $\mu = x,y,z$, is
 \begin{equation}
 \epsilon_{\ell, \ell'}( \vec k,T) \equiv \{ g_L(T) + \xi_0^2 {\vec K} ^2 \} \delta_{\ell,\ell'}+ \frac{ A_1}{2} U_{\ell\ell'}(\hat k). 
\end{equation}

The anisotropic Compatibility  kernel  in the  energy spectrum  can induce preferred DW orientations \cite{R14,R15,R16,R21}. For example the $\ell = \ell'$  kernel  $U_{\ell,\ell} (\hat k)$ is smallest $U_{\ell,\ell}(min) =0$  at the  most favoured  orientation, and largest  $U_{\ell,\ell}(max) >0$ for most disfavoured.  The  negative sign of the Landau term $H_L  \sim g_L <0$ and the positive signs of the  Ginzburg  term $H_G \sim \vec k^2 > 0$ and the Compatibility term $ H_C > 0$ imply the spectrum  $\epsilon_{\ell, \ell} (\vec k,T)$ could vanish along some Fourier contour. This contour will be angularly modulated, through  the anisotropy  of the Compatibility kernel \cite{R15,R16}.

 In MC simulations, the initial state $t=0$ is high-temperature austenite that is  randomly and dilutely ($2 \%$) seeded with martensite unit-cells.    Typical parameters are $T_0=1$;  $\xi_0 ^2=1$; $T_c=0.95$; $E_0=3$; system volume $N=L^3=16^3$; $N_{runs}=100$; and holding times $t_h=10^4$ MC sweeps.The martensite fraction is $n_m(t) \equiv \frac{1}{N} \sum_{\vec r} { S}^2(\vec r, t) \le 1$, with $n_m=0$ or $1$ for uniform austenite or martensite.  The  conversion time $t_m$  is defined as when \cite{R16}   $n_m (t_m) =1/2$. An athermal martensite droplet or embryo  can  rapidly form  anywhere, and after waiting till $t_w=t_m$, can  {\it propagate}  rapidly to the rest of the system \cite{R13}. Hence  it is mean  {\it rates} ${\bar r}_m$ (or inverse times),  that are averaged over runs, analogous to total  resistors in parallel  determined by the smallest resistance. Mean times ${\bar t}_m$ are inverse mean rates:  ${\bar t}_m (T) \equiv 1/{{\bar r}_m (T)}$.

 \begin{figure}[h]
\begin{center}
\includegraphics[height=6.5cm, width=8.0cm]{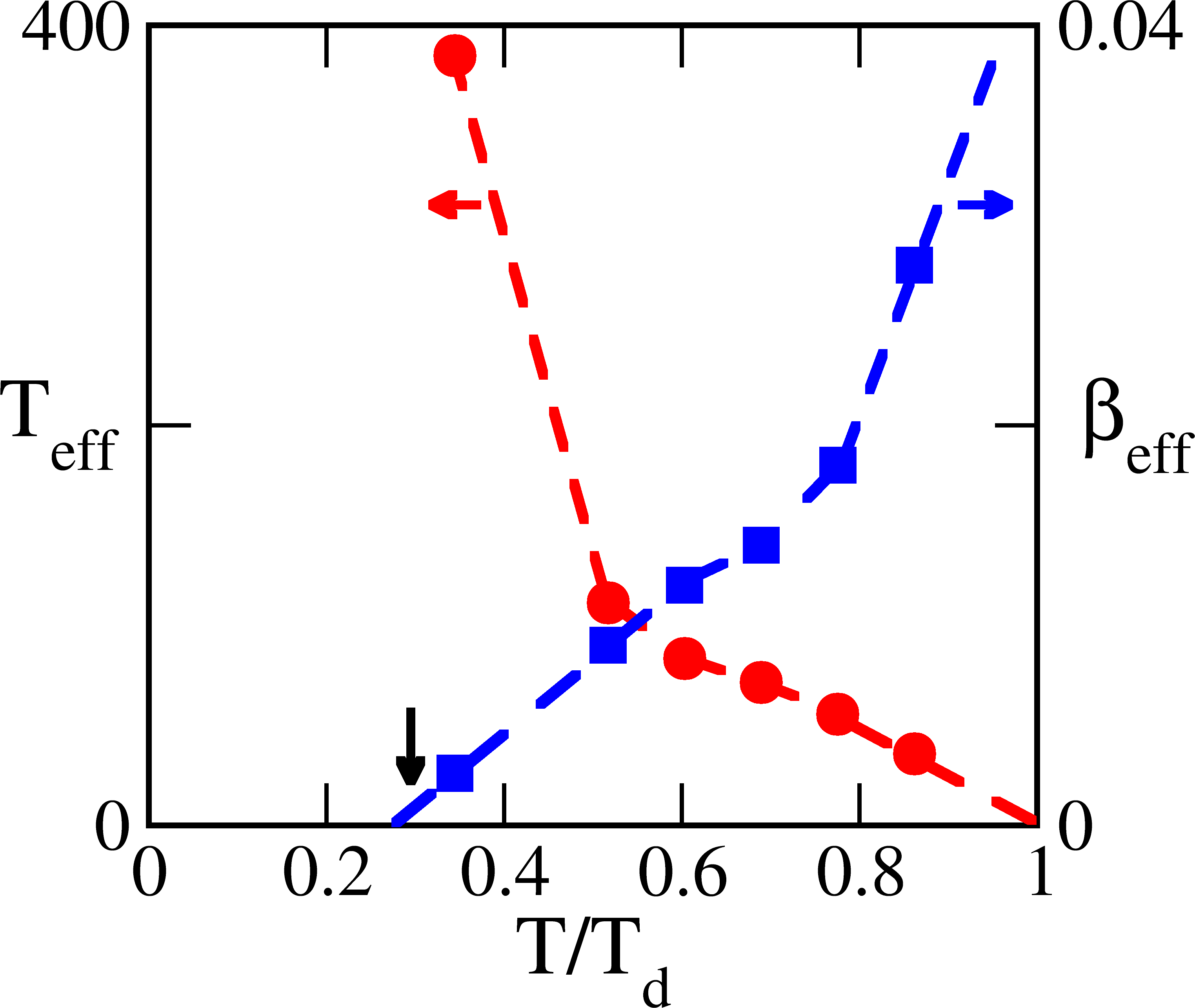}
\caption{{\it Effective temperature and  its inverse, versus quench temperatures for CT transition: }   Left  vertical axis: $T_{eff}(T)$  versus $ T/T_d$ appears to  vanish as $\sim(T_d-T)$, and rises from $T_1 < T_d$. Right vertical axis: $\beta_{eff} (T) $  appears to vanish as $\sim (T-T_1)$,  and rises  rapidly towards $T_d > T_1$.  The entropy barrier $S_B \sim \beta_{eff}$ will then vanish at $T_1$ (downward arrow) or diverge  at $T_d$. Dashed lines are guides to the eye.}
\end{center}
\end{figure}

 The MC procedure is standard, but with a crucial extra data retention \cite{R6,R7,R8,R9} of energy changes. \\
0. Take $N$ sites, each with a  vector spin of $N_{OP}$ components,  in one of $N_V +1$ possible values (including zero) at MC time $t$. Each $\{{\vec S}(\vec r)\}$ set is a `configuration'.  \\
1. Randomly pick one of  $N$ sites, and randomly flip the spin  on it  to a {\it new} direction/value, and find  the (positive/negative) $\delta E$ changes for the new configuration.\\
2. If the energy change $\delta E \leq 0$, then accept the flip. If $\delta E > 0$, then accept flip with probability $e^{- \delta E /T }$. {\it Record} this $\delta E$, that is not usually retained after use.\\
3. Repeat steps 1 and 2. Stop after $N$ such spin-flips. This configuration has the conversion fraction $n_m (t+1)$. \\
4. We collect  \cite{R24} all $ \{ \delta E\}$ from each spin-flip (configuration change) within each MC sweep of every run,  up to the conversion time for that run, $t \leq t_m (T) \leq t_h$.
 The set size $N \times { t}_m \times N_{run}$ has up to  $16^3 \times 10^4 \times 100 $ data points. We take six quenches, from $T=T_1$ up  to $T_d$. 

 Figure 2a shows $n_m(t)$,  the martensite conversion-fraction in a  {\it single} run, versus MC time $t$ for different  temperatures $T$. For  quenches $T\leq T_1$,  avalanche  conversions, characteristic of athermal martensite, occur in the very  first sweep over all spins  ($t = 1$).  We identify $T_1$ with the martensite start temperature \cite{R11,R12}      $M_s=T_1$.  For higher temperatures $T > T_1$,  there is a curious `incubation' period,  when nothing happens macroscopically,  until a postponed avalanche at  $t_w=t_m$. These models\cite{R16,R24} display the delayed transitions  and burst-like growth of order, characteristic of martensites  and manganites \cite{R11,R12,R17}. Fig 2b shows that  for $T$ above $T_1$ (downward arrow), and approaching $T_d$, the mean  incubation delays rise steeply, due to entropic bottlenecks.

\begin{figure}[ht]
\begin{center}
\includegraphics[height=6.5cm, width=8.5cm]{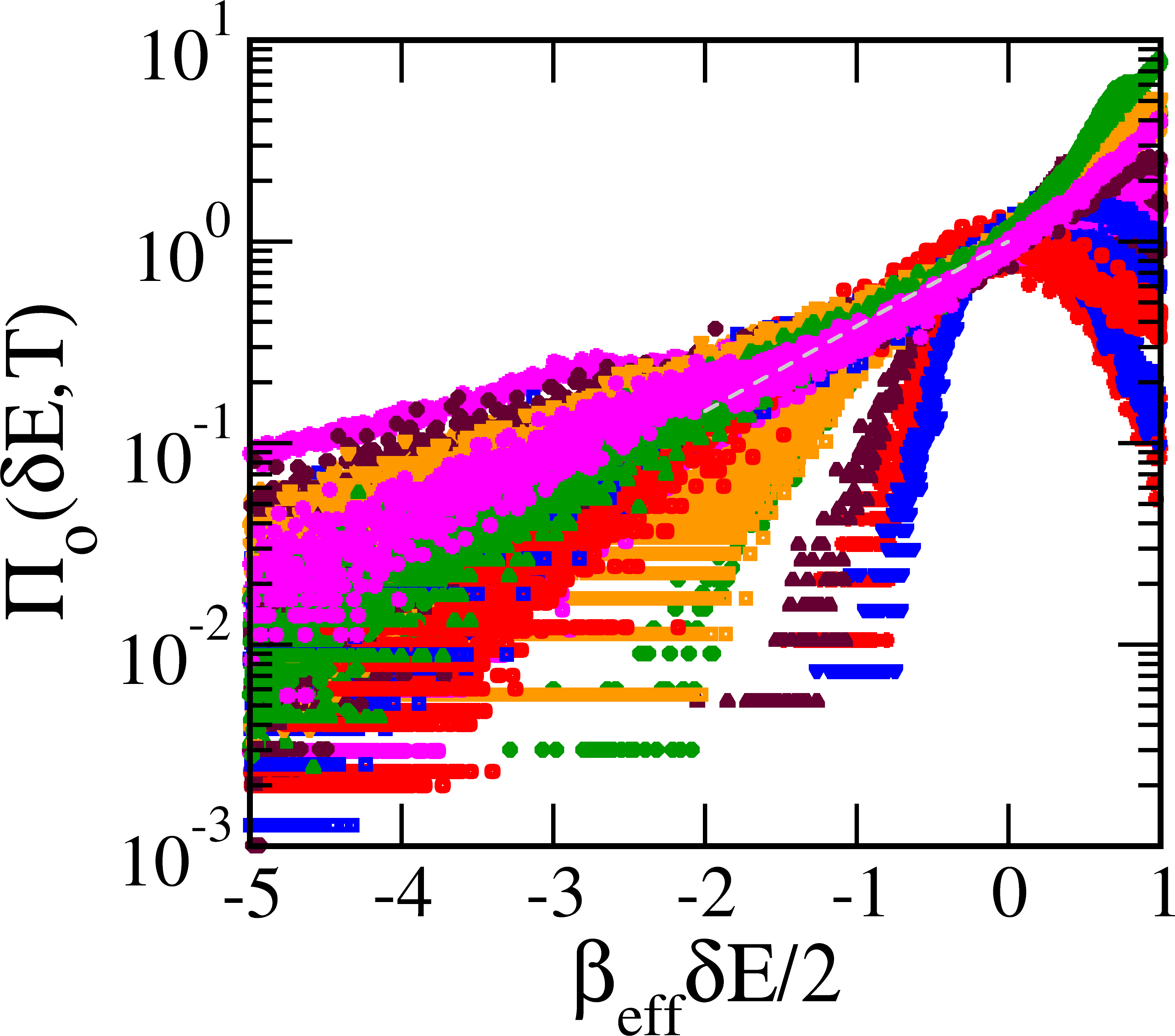}
\caption{{\it Universal slope of PES distribution:} Log-linear scaled  plot of  $\Pi_0 (\delta E,T) \equiv P_0(\delta E,T)/ P_0(0,T)$ versus  $z \equiv \beta_{eff} (T) \delta E/2$. The PES  predicts a universal slope of unity at $z=0$. The four transitions mentioned have respective slope averages and standard deviations of $1.000\pm 0.045, 1.025 \pm 0.036, 1.009 \pm 0.08, 0.850\pm 0.085$. The data for six $T$ and four transitions have mean slope (dashed  white line)  of $0.97 \pm 0.06$. } 
 \end{center}
\end{figure}

\begin{figure}[h]
\begin{center}                   
\includegraphics[height=4.0cm, width=8.5cm]{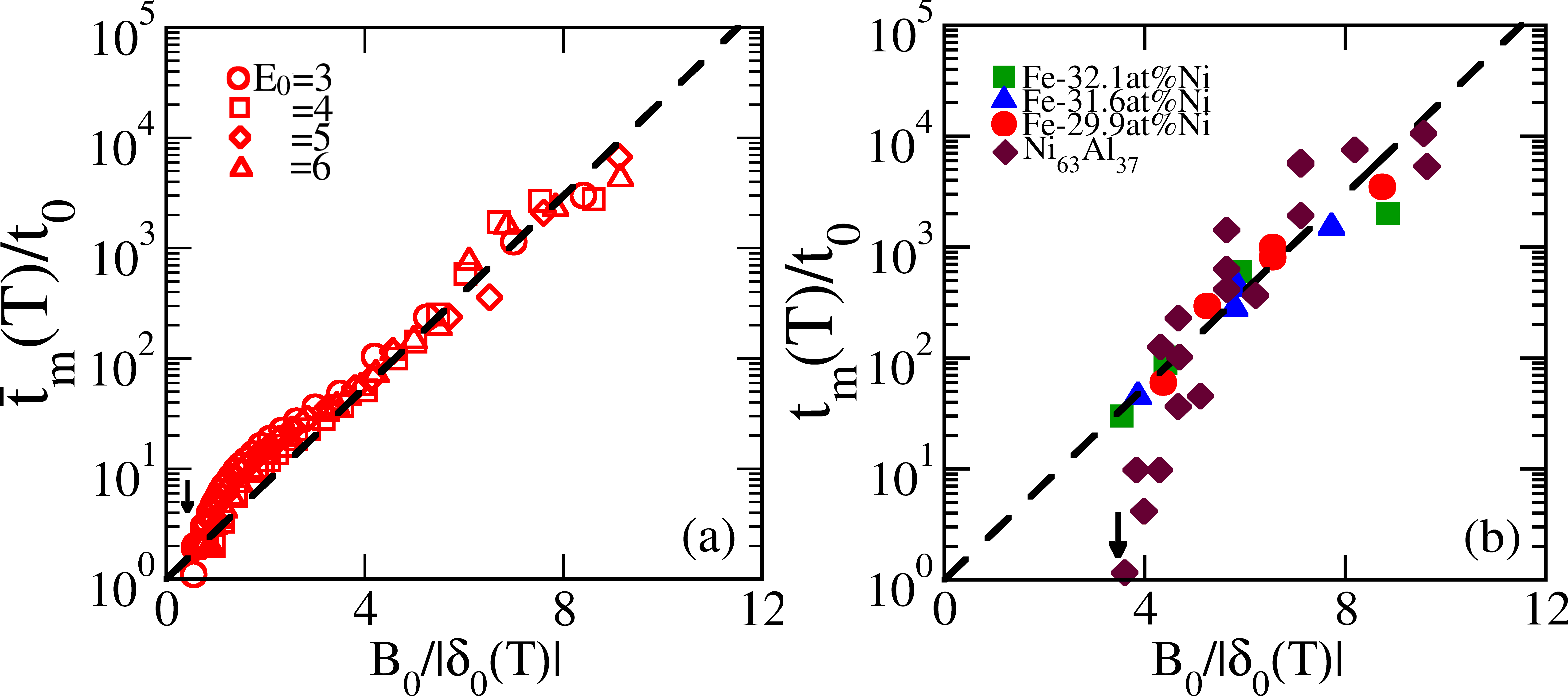}
\caption{{\it Log-linear plots of scaled  time versus (inverse) scaled  temperature deviation: } Scaled  conversion times $t_m/t_0$  versus $B_0/|\delta_0|$ with parameters $t_0,B_0$ extracted from data. There is Vogel Fulcher linearity near $T_d$, with falloffs near $T_1$ (downward arrows).   a) From  CT simulations in 3D  for  different $E_0$. b) From experiments \cite{R11,R12} in  3D for  different alloys. }
\end{center}
\end{figure}

For our model,  the boundary of a 3D bottleneck is from  the spectrum  set equal to zero $\epsilon_{\ell,\ell} (\vec k) =0$,  defining an {\it anisotropic} surface in $\vec k$-space. For the CT case, a [1,1,1] slice can  intersect the  bottleneck surface  as an open,  butterfly-shaped locus with an inner and outer radius, inside the Brillouin Zone (BZ). See Fig 3. For $T \leq T_1$,  the  radius $k_{out} (T)$ is larger than a BZ scale  $\sim \pi$, and martensitic passages are immediate.  For $T_1 < T < T_d <T_0, $ the butterfly bottleneck shrinks on warming. At $T=T_d$, the inner squared-radius $k_{in} (T)^2 = |g_L (T)| - (A_1/2) U_{\ell,\ell} (max)$ vanishes, and the topology of the connected  butterfly changes to that of   a  segmented four-petaled flower: entropy barriers diverge, and PES heat releases are arrested. For the `precursor' region \cite{R14} $T_d < T < T_0$, PES passages are energetically available, but  entropically inaccessisible. Repeated bottleneck entry attempts could induce vibrations. See Video\cite{R21} D. Finally, at $T=T_0 =1$, the outer squared-radius $k_{out} (T)^ 2 = |g_L (T)|$ also vanishes: the bottleneck becomes a point, and  only austenite exists.

We collect the  $O(1)$ changes $ \{\delta E \}$  to the $O(N)$ energy $E$.
The probability $P_0 (\delta E,T)$ to   access $E'$ from $E$,   is proportional to the number of target states $ \Omega  (E')$.  With $S(E') = \ln \Omega (E') $, the probability ratio $R_0 (\delta E)$   of energy changes  is related to the entropy change  $\Delta S (\delta E) \equiv S(E')-S(E)  < 0$  by a fluctuation relation for aging \cite{R1,R2,R3,R4} :

\begin{equation}
R_0 \equiv \frac{P_0 (\delta E,T)}{ P_0 (-\delta E,T)}  = \frac{\Omega (E') }{\Omega (E)}  = e^{\Delta S (\delta E)}.
\end{equation} 
Entropy barriers $S_B \equiv -\Delta S > 0$  rise, when the searched-for states become rarer.   
Since $R_0(\delta E) R_0 (-\delta E) \equiv 1$, the  entropy change is odd, $\Delta S(\delta E) +\Delta S(-\delta E)=0$, and  a solution for the PES distribution is

\begin{equation}
P_0 (\delta E,T)=  {P_0}^{(+)} (\delta E) ~  e^{\frac{1}{2} \Delta S (\delta E)},
\end{equation} 
 with  an even ${P_0}^{(+)} (\delta E) \equiv  \sqrt{P_0(\delta E,T) ~P_0(-\delta E,T)}$. The leading  entropy-change term  for small heat releases is  $\Delta S \simeq \beta_{eff} \delta E$ where $\beta_{eff} \equiv 1/ T_{eff}$. For $\delta E = - |\delta E| <0$,  the  Boltzmann-like form $P_0 \simeq e^{-\frac{1}{2} \beta_{eff}(T) |\delta E|}$ gives a physical meaning to the effective temperature, as a {\it search  range} denoting accessible energy shells. If  $\beta_{eff} \rightarrow 0$, entropy barriers collapse, and  passages are immediate.  If $T_{eff} \rightarrow 0$, then entropy barriers diverge, and  passages cease. Glass-like freezing is a shutdown of PES searches.

Fig 4a  shows that,  as in PES models \cite{R6,R7,R8,R9}, the $P_0 (\delta E,T)$  peaks are  at positive $\delta E$, understood as a completion-of-square between a gaussian peaked at the origin  and an exponential tail for $\delta E < 0$. Fig 4b shows a zoom-in near  the origin, where the  slopes define  $\frac{1}{2} \beta_{eff} (T)$.

Fig 5 shows the dependence of  $\beta_{eff} (T)$ and  $T_{eff} (T)$   on the quench temperature  $T$. The data suggest a linear vanishing of  $ T_{eff} \simeq  (T_d-T)/ (B_0 T_d)$  near $T_d$, at a  search freezing and a suppression of the heat releases to the bath. There is also a linear vanishing of $\beta_{eff} \sim T-T_1$  near $T_1$, at a search avalanche and prompt equilibration.

Fig 6 shows log-linear plots  for a scaled  $ \Pi_0  (\delta E,T)\equiv P_0 (\delta E,T) /P_0 (0,T)$ versus the entropy-barrier related  variable $\frac{1}{2} \beta_{eff} \delta E$. Data are for four 3D structural transitions \cite{R18,R23}, and six $T$  between the collapse ($T_1$)  and divergence ($T_d$)  of entropy barriers.

The mean conversion time  is exponential in the entropy barrier \cite{R23} ${\bar t} \sim e^{1/T_{eff} (T)}$, so near $T_d$ we have   ${\bar t}_m (T) \simeq  t_0 e^{B_0/|\delta_0 (T)|}$, where the constants $B_0,t_0$ can be fixed by simulational and experimental data \cite{R24}.  The initial slope  in  $|\delta_0|$ of
 $1/\ln{ {\bar t}_m (T)}$  gives $1/B_0$; and the extrapolated intercept of $\ln{ {\bar t}_m (T)}$  versus $B_0/ |\delta_0 (T_0)|$ gives $t_0$. For   Ni-Al data \cite{R11} the `fragility' parameter \cite{R2} $B_0 T_d \simeq 1.23 $ Kelvin, and  the austenite-martensite DW hop time is $t_0 \simeq 1 $ sec.

 Fig 7a shows that CT times show VFT  behaviour near $T_d$ and fall-off behaviour near $T_1$. Fig 7b shows data extracted from  Ni-Al and Fe-Al alloys\cite{R11,R12} are similar.  

Signatures of PES could be sought, in previous  simulations or experiments\cite{R3,R5} under systematic temperature quenches, with a recording of energy releases.
 Further experimental work on martensitic alloys \cite{R11,R12}  could  record signal and noise under systematic quench steps of $1/|\delta_0|$, over the delay region $T_d > T> T_1$; as well as the precursor \cite{R14,R21} region $T_0 > T> T_d$ above it.  Non-stationary distributions of   energy releases might  be determined  through concurrent resistive, photonic, acoustic, and elastic signals\cite{R25}.  Finally,  one might speculate that complex oxides quenched near their  structural/ functional  transitions, could show  PES ageing behaviour in their (strain-coupled)  functional variables\cite{R17,R18}. \\

 In summary, post-quench  ageing in  athermal martensites shows characteristic signatures of the Partial Equilibration Scenario. The conversion arrest and delay-divergence found in 3D simulations and alloy experiments, are understood as arising from a  vanishing  of the search temperature that governs the PES cooling process.
 
 Acknowledgement: It is a pleasure to thank Smarajit Karmakar for valuable discussions on the glass transition.


\begin{thebibliography}{2004}
\bibitem{R18}
The four chosen structural  transitions can occur in materials with useful functionalities \cite{R17}. The transitions are:  tetragonal-orthorhombic ($YBCO$, superconductivity); cubic-tetragonal ($FePd$, shape memory);  cubic-orthorhombic ($BaTiO$, ferroelectrics); cubic-trigonal ($LaSrMnO$, colossal magnetoresistance).
\bibitem{R23}
The mean equilibration rate is  the average over an inter-shell rate $1/t(\delta E)$ over all heat releases. With a normalized PES probability, 
$1/{\bar t} ={ \int_{-\infty}^{0} d\delta E~ P_0 (\delta E, T)/ t(\delta E)}$. For slow intershell variation $1/t(\delta E) \simeq 1/t_0$, the mean rate is proportional to the acceptance fraction over energy decrements. Taking the  PES probability, as a shifted gaussian with peak mean  energy change $M >0$, and variance $\sigma^2 = 2M T_{eff}$, the  acceptance fraction is a complementary error function, whose asymptotic behaviour yields $t_0 /{\bar t} \simeq e^{-M/4T_{eff}}$.  Dropping constants, the martensitic mean conversion time near $T_d$ is written 
simply as ${\bar t}_m (T) = t_0 ~ e^{1/T_{eff}(T)}$.
\end{thebibliography}
\end{document}